\documentclass[epj,final]{svjour}
\usepackage{graphics}
\begin{document}
\title{QCD Saturation and Soft Processes}
\author{Errol Gotsman\ } 
%
%
\institute{School of Physics and Astronomy, Tel Aviv University,
Ramat Aviv 69978, Israel }
\date{Received: date / Revised version: date}
%
\abstract{
 We show that an approximate solution to the amended non-linear Balitsky-Kovchegov 
evolution equation which was formulated for hard (large $Q^{2}$) QCD processes,
can be  extended  to provide a good description of photoproduction and
 soft hadronic (non perturbative) 
reactions.
} 
\maketitle
\section{Introduction}
\label{intro}
 Further proof that the linear QCD evolution equations fail to describe the DIS data
for low values of $Q^{2}$ has recently been provided by the ZEUS collaboration 
\cite{Re1}. See Fig.1. For values of $Q^{2} \; \geq \; 1 \; GeV^{2}$ the ZEUS NLO QCD fit 
provides an excellent description of the data, however, as $Q^{2}$ becomes smaller
the discrepancy between the predictions and the data increases.
\begin{figure}
\resizebox{0.7\hsize}{!}{\includegraphics{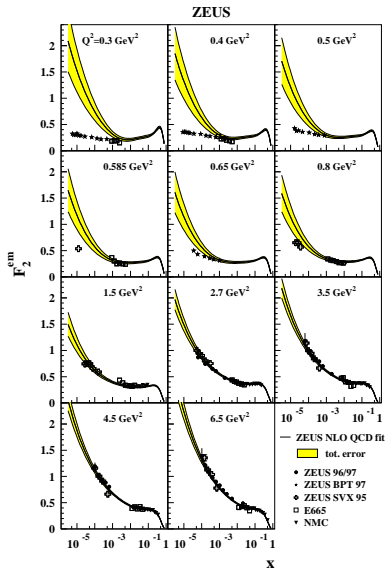}}
\caption{$F_{2}$ data at very low $Q^{2}$ compared to ZEUS-S NLO QCD fit, from ref.[1].}
\label{fig:1}       
\end{figure}
  
   The reason for this is well known, and is due to the fact  that 
in the linear evolution equations DGLAP and BFKL, the splitting functions only 
incorporate the production of partons (gluons). GLR \cite{GLR} in their classical paper
pointed out that when the density of partons becomes sufficiently large (that they 
overlap), one has to include non-linear (annihilation) processes in the  evolution
equations.
A recent application of the ideas of GLR using the dipole formalism has been suggested
by Balitsky \cite{Bal} and Kovchegov \cite{Kov} (which we will denote by BK). The 
advantage of the BK equation is that: \\
(i) it accounts for saturation effects due to high parton densities; \\
(ii) it sums higher twist contributions; \\
(iii) it allows one to extrapolate to small values of $Q^{2}$ (large distances).\\

\section{Amended Non-linear BK Equation}
\label{sec:1}
%
 The BK equation can be written in the form \cite{GLLM}
\begin{eqnarray}
\frac{d \tilde{N}({\mathbf{x_{01}}},b,Y)}{d Y}\,\,\,&=&\,\,
\,\,\,\frac{C_F\,\alpha_{s}}{\pi^2}\,\,
\int_{\rho} \,\,d^2 {\mathbf{x_{2}}}\,
\frac{{\mathbf{x^2_{01}}}}{{\mathbf{x^2_{02}}}\,
{\mathbf{x^2_{12}}}}\,\nonumber \\
\cdot\,\,\,
(\,\,2\,\tilde{N}({\mathbf{x_{02}}},b,Y)  
 &-&  \,\,\,\tilde{N}({\mathbf{x_{02}}},b,Y)
\,\,\tilde{N}({\mathbf{x_{12}}},b,Y)\,\,) 
\end{eqnarray}
where $\tilde{N}({\mathbf{r_{\perp}}},b,Y)$ denotes the imaginary part
of the  amplitude  of a dipole of size $ r_{\perp}$ elastically scattered at impact 
parameter b,   i.e.
\begin{equation} 
 \tilde{N}({\mathbf{r_{\perp}}},b,Y) \; \; = \;\; 
Im a^{el}_{dipole}({\mathbf{r_{\perp}}},b,Y).
\end{equation}  
$Y  =  ln(1/x_{BJ})$ denotes the rapidity. In Eq. (1) the linear term 
corresponds to the 
 LO BFKL evolution, while the non-linear negative term is responsible for unitarization.
\par
   A deficiency of the BK equation is that it does not contain evolution in $Q^{2}$,
and therefore lacks the correct short distance behaviour. To remedy this we have 
introduced a correcting function $\Delta \tilde{N}({\mathbf{r_{\perp}}},b,Y) $
which accounts for the DGLAP evolution in $Q^{2}$.
  Our full solution  therefore consists of the sum of two terms 
\begin{equation}
  N({\mathbf{r_{\perp}}},b,Y) \; \; = \; \;  \tilde{N}({\mathbf{r_{\perp}}},b,Y) \; 
+ \; \Delta \tilde{N}({\mathbf{r_{\perp}}},b,Y) 
\end{equation}
Full details of the calculation are contained in ref. \cite{GLLM}, where we have 
simplified the treatment of the dependence on the impact parameter b. We solve for the 
initial condition at b = 0, and then restore b dependence by assuming factorization
and that the dipole profile function inside the target is given by a Gaussian 
distribution
$S(b)  =  e^{-\frac{b^{2}}{R^{2}}} $ for all b. This is obviously an over 
simplification (see discussion in Section 5), and  the  true  impact 
parameter dependence of $N({\mathbf{r_{\perp}}},b,Y)$ is far more complex.

\par
   The deep inelastic structure function $F_{2}$ is related to the dipole cross section
(defined in Eq.(2))
\begin{eqnarray}
F_{2}(x_{BJ},Q^{2}) \;\; &=&\; \; \frac{Q^{2}}{4 \pi^{2}} \int d^{2}r_{\perp} 
\int dz P^{\gamma^{*}}(Q^{2};r_{\perp},z) \nonumber   \\
\;\;&\cdot& \;\;  \sigma_{dipole}(r_{\perp},x_{BJ})
\end{eqnarray}
Where  $P^{\gamma^{*}}(Q^{2};r_{\perp},z)$     denotes the probability of the decay of a 
virtual photon having four momentum $Q^{2}$ into a colourless dipole ($q\bar{q}$ pair) of 
size $r_{\perp}$, with the quark (anti-quark) taking a fraction z (1-z) of the virtual 
photon's momentum.

\section{Numerical solution of the equation for DIS}
\label{sec:2}
   In making our fit we include all available data satisfying the following criteria
$ 10^{-7} \; \leq \; x_{BJ} \; \leq \;   x_{0 BJ} \; = \; 10^{-2} $ and 
$ Q^{2} \; \geq \; 0.04\; GeV^{2}$. We solve Eq.(1) as an evolution equation in rapidity 
with a fixed grid in $r_{\perp} \; ( = 2/Q $ GeV)  space and a dynamical step in 
rapidity.
We fit to 345 data points for $F_{2}(x_{BJ},Q^{2})$, 
  and obtain an overall  $\chi^{2}/df \; \approx \;1$. In ref. \cite{GLLM} we also
compare to data on the logarithmic slopes of $F_{2}$ i.e. 
$\frac{d F_{2}}{d ln Q^{2}}$ and $\frac{d ln F_{2}}{d (ln 1/x)}$.
 The values that we obtained  for
$\lambda \; = \; \frac{d ln F_{2}}{d (ln 1/x)}$ were
 $\lambda \approx 0.07$ for $Q^{2} = 0.05 \; GeV^{2}$ and $x_{BJ} = 10^{-5}$, 
increasing to
$\lambda \approx 0.3$ for $Q^{2} = 150 \; GeV^{2}$ and $x_{BJ} = 4. 10^{-2}$. i.e.
our formalism was successful in describing not only the short range  (large
$Q^{2}$) data but also the "soft" data ( at very low $Q^{2}$ ) where the traditional 
Pomeron 
intercept is $\approx \; 0.08$.
 
\section{Extension of BK formalism to photoproduction}
\label{sec:3}

\par
   The surprising results discussed above, that with the \\
 amended BK equation
we found an excellent description of all DIS data for $0.05 \; \leq Q^{2} \leq 200 
 \; GeV^{2}$  prompted the question, whether this formalism could also successfully
describe soft processes e.g. photoproduction?

    To extend our formalism to photoproduction \cite{Bar1}, it is necessary to make the 
following 
alterations: \\
(i) We need to introduce a finite mass as a cutoff for the $r_{\perp}$ integration
in Eq.(4 ),  we take this parameter as $m_{q}$ (the quark mass). \\
(ii)   The variable $x \;(=x_{BJ})$ is not defined for $\gamma$-p scattering,
and we relate $x$ to the energy by introducing a non-perturbative scale $Q^{2}_{0}$,
 and taking $x \; = \; \frac{Q^{2} \; + \; Q^{2}_{0}}{W^{2}}$.
To reduce the number of free parameters we have set 
$Q^{2}_{0} \; = \; 4\; m_{q}^{2}$. On fitting to the high energy photoproduction data
we find a value of $m_{q}\; =  0.15 $  GeV.  
    In the colour dipole formalism one can only hope to reproduce the asymptotic energy 
dependence i.e. the Pomeron contribution. We also need to include a non-singlet (NS) 
term to account for lower energy (higher $x$) contributions.
We test two possible forms for the NS term: \\
1) Based on Valence Quark Model (VQ):
\begin{eqnarray}
 F^{VQ}_{2} \; = \; ( \frac{Q^{2}}{1 \; GeV^{2}})^{(1 +\beta)}
 \cdot  \frac{( 1 \; GeV^{2} \; + \; \mu^{2})}{(Q^{2} \; + \; \mu^{2})} \nonumber \\
  \cdot \sum_{i=u,d}e^{2}_{i}q^{V}_{i}(Q^{2}  = \;  1 \; GeV^{2}) 
\end{eqnarray}
We freeze the CTEQ6 valence quark contributions at $Q^{2} \; = \; 1 \; GeV^{2}$,
and on fitting to data, the best fit parameters are  $\beta \; = \; 0$ and 
$\mu^{2} \; =\; 0.13 \; GeV^{2}$.

2) Based on exchange of Secondary Reggeons (SR):
$$ \sigma^{\gamma p}_{SR} \; = \; f(0) \frac{\tilde{M}^{2}}{(Q^{2} \; + \; 
\tilde{M}^{2})}
\cdot (\frac{s}{s_{0}})^{\alpha_{R}} $$
with $\alpha_{R}  =  - 0.45$ and $s_{0}  = \; 1 \:  GeV^{2}$, $f(0)$
denotes the residue at $Q^{2}  = 0$. Fit results in $f(0)  = 0.19$ mb and
$\tilde{M}^{2}  =  2 \; GeV^{2}$.

    The results of the fit to photoproduction and low $Q^{2}$ DIS data
with the two alternate low energy contributions are shown 
in Fig.2.

\begin{figure}
\resizebox{0.7\hsize}{!}{\includegraphics*{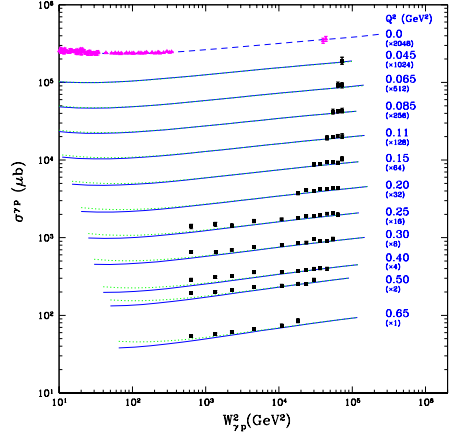}}
\caption{$\gamma$-p and DIS cross-sections at very low $Q^{2}$. Solid line VQ and dotted
line SR parametrization for low energy contribution (see text for details). }
\label{fig:2}       
\end{figure}

\section{ The BK formalism and soft hadronic processes}
\label{sec:4}

      Based on the  successful extension of our BK formalism to photoproduction (Sec.4), 
we also 
applied the  procedure  to soft hadronic interactions \cite{Bar2}. 
    This necessitated adapting the basic formula Eq.(4) of the dipole model, and 
hypothesizing that the hadron-proton cross-section is given by:
\begin{equation}
\sigma_{hadron-proton}(x) \; = \; \int d^{2}r_{\perp} \mid \psi_{h}(r_{\perp}) \mid^{2}
\sigma_{dipole}(r_{\perp},x)
\end{equation} 
where $\psi_{h}(r_{\perp})$ represents the wave function of the hadron which scatters off 
the  target proton, and  the energy dependence is given by
  $x = \frac{Q_{0}^{2}}{W^{2}}$, with $Q^{2}_{0}$ being an additional non-perturbative 
scale.

 For the hadronic wave functions we use the form suggested by the Heidelberg group,
 Dosch et al. \cite{Dosch}.
The hadronic transverse wave function is taken as a simple Gaussian, where the square of 
the wave function is given by
\begin{equation}
\mid \psi_{h}(r_{\perp}) \mid^{2} = \frac{1}{\pi S^{2}_{M}} exp(- 
\frac{r_{\perp}}{S^{2}_{M}})
\end{equation}
and $S_{M}$ is a parameter related to the meson size.
We have used $S_{\pi}$ = 1.08 fm and $S_{K}$ = 0.95 fm, which is  related to the 
electromagnetic radii. The exotic channel $K^{+}$ p has no secondary Regge contributions,
while for the $\pi^{-}$ p reaction we have added a secondary contribution \'{a} la 
Donnachie and Landshoff \cite{DL}. Our results are displayed in Fig.3,
 for more details regarding the fit shown in Fig.3 we refer the reader to 
\cite{Bar2}.

\begin{figure}
\resizebox{0.6\hsize}{!}{\includegraphics{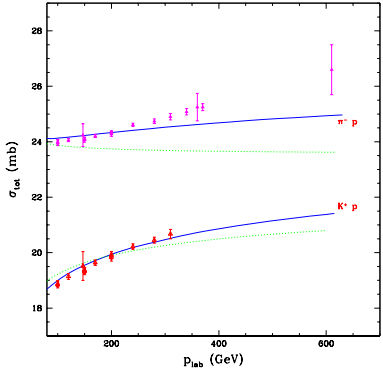}}
\caption{ $K^{+}$p and $\pi^{-}$p cross-sections. The full line is the prediction in 
our model and the dotted line using the Golec-Biernat W$\ddot{u}$sthoff dipole 
parametrization \cite{GBW} }
\label{fig:3}       
\end{figure}

 For  a baryon projectile, we assume that the baryon is constituted
of two colour dipoles, one dipole formed around two quarks, and the second dipole from 
the centre of mass of these two quarks to the third quark in the baryon. Our results for 
$\bar{p}$-p  and p-p scattering are shown in Fig.4. The assumed Gaussian dependence in b, 
corresponds to a  
 $e^{-t}$ behaviour of the differential cross-section. The full lines in Fig.4 show 
that this is a poor description, a much better fit is obtained by assuming that the
momentum transfer dependence is of a "form factor" dipole type, which transforms to a
Bessel function 
$K_{1}$ dependence in impact parameter space. See ref. \cite{Bar2} for more details. 

\begin{figure}
\resizebox{0.6\hsize}{!}{\includegraphics{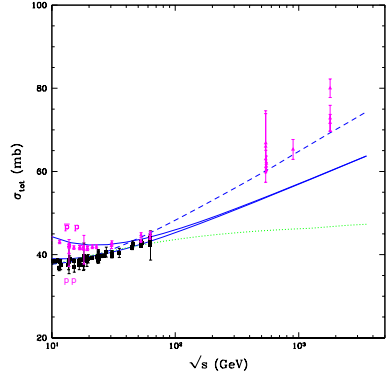}}
\caption{ $\bar{p}$p and pp total cross-sections. The full lines are the model predictions
with a Gaussian profile for the impact parameter dependence. The dashed line for a 
$K_{1}$ profile and dotted line for the GBW dipole model.
}
\label{fig:4}       
\end{figure}

\section{ Conclusions}
\label{sec:6}

    We have shown that our approximate solution to the amended non-linear BK equation
which was formulated for hard (large $Q^{2}$) processes \cite{GLLM},
can be successfully extended to describe photoproduction and soft (non perturbative) 
hadronic reactions.  
  Two  outstanding problems remain:\\
(i)  the impact parameter
dependence of the colour dipole. The assumption of factorization in b space, plus 
imposing a Gaussian like behaviour in b (for all values of b), is obviously naive,
as can be judged from the results we have obtained for $\bar{p}p$ scattering. The
search for the correct impact parameter dependence of the solution to the BK equations
continues. \\
(ii) the form of the hadronic  QCD wave function. The form suggested by \cite{Dosch},
is only a first approximation and should be improved. 

   I would like to thank my colleagues Jochen Bartels, Genya Levin, Michael Lublinsky 
and Uri Maor for a most enjoyable collaboration, the fruits of which are presented
here.
%

\end{document}